\documentstyle{aipproc}
\begin{document}
\title{Meson Spectroscopy and the \\ Search for Exotics}

\author{Curtis A. Meyer}
\address{Carnegie Mellon University, Pittsburgh, Pennsylvania 15213}

\maketitle

\begin{abstract}
I will review the current status of exotic hadrons in the meson sector.
There is currently strong evidence that a scalar glueball mixed into
the normal scalar mesons has been found. There is also an interesting
candidate for the tensor glueball state. Finally, I will discuss
hybrid mesons with explicitly exotic quantum numbers.
\end{abstract}
\section*{Introduction}
In the context of this paper I will review the current status of
exotic mesons; glueballs and hybrid mesons. Our current best
evidence for a glueball comes from the scalar meson sector, $J^{PC}=0^{++}$.
In this sector there appears to be two states, the $f_{0}(1500)$ and
$f_{0}(1750)$ which are produced in several glue--rich mechanisms.
The decay rates of the $f_{0}(1500)$ have been studied in detail by
the Crystal Barrel experiment which finds that this state
cannot be explained as either a pure glueball nor a pure meson.
The $f_{0}(1500)$ combined with the $f_{0}(1750)$ leads us to the 
interpretation that these states are mixtures of both glueball and 
normal mesons. In the tensor sector, there is currently a very interesting 
candidate state, the $\xi(2230)$. Its best evidence comes from
radiative $J/\psi$ decays, and appears to have several properties
suggestive of glueballs. Lastly, there is evidence from E852 at 
Brookhaven for a state with explicitly exotic quantum numbers.
However several properties of this state seem to disagree 
with expectations for a hybrid meson.
\section*{Glueballs}
\subsection*{The Theoretical Situation}
A pure glueball is a bound state of only gluons. Due to the 
fact that gluons carry the color charges of QCD, it is theoretically 
possible for them to form bound states devoid of any quark content. 
The quantum numbers of these states are derived by considering them 
to contain either 2 or 3 {\em valence} gluons. While we are still
unable to solve QCD exactly in the nonperturbative regime, most models
which are able to explain observed phenomena predict that glueballs
should exist, and most predict the lightest will be the scalar,
$J^{PC}=0^{++}$. At the moment, we believe that lattice calculations
come closest to actually solving non--perturbative QCD. A calculation
of the entire glueball spectrum~\cite{uk-glueball} finds that the
scalar glueball has a mass of $1550\pm 50$MeV/c$^{2}$ while the next
lightest state is the tensor at a mass of $2270\pm 100$MeV/c$^{2}$.
A recent calculation with a factor of 10 improvement in lattice
density predicts that the scalar glueball has a mass of 
$1740\pm 71$MeV/c$^{2}$, and for the first time makes predictions
for decay rates into flavorless pseudoscalar meson pairs as a 
function of mass~\cite{ibm-glueball}. Given that the two groups 
have a different procedures to extrapolate to the continuum limit, 
we take the average of these, $1610\pm70\pm130$ MeV/c$^{2}$,
as the prediction for the mass of a pure scalar glueball.

Next we consider where we should search for glueballs. There
are several reactions which are considered as glue--rich. Radiative 
$J/\psi$ decays, $\psi\rightarrow\gamma X$ are generally considered
the best source of glueballs simply because the $c$ and $\bar{c}$ quark
have to annihilate in order for the decay to proceed. Defining 
$b(R_{J}\rightarrow xx)=\Gamma(R_{J}\rightarrow xx)/\Gamma_{\rm tot}$,
expectations from~\cite{zhenping} give:
\begin{eqnarray}
b(R(\bar{q}q)\rightarrow gg) & \simeq & 0.1 \sim 0.2 \label{eq:normal} \\
b(R(G)\rightarrow gg) & \simeq & 0.5 \sim 1 \label{eq:glue}.
\end{eqnarray}
Where equation~\ref{eq:normal} is for a normal meson and 
equation~\ref{eq:glue} is for a glueball. These quantities can be 
related to the radiative decay rate via equations~\ref{0++-rate} 
and~\ref{2++-rate}.
\begin{eqnarray}
(10^{3}) B(J/\psi \rightarrow \gamma R(0^{++})) & = & 
\left(\frac{m}{1500 MeV}\right) 
\left(\frac{\Gamma_{R\rightarrow gg}}{96 MeV}\right ) 
\frac{x\mid H_{T} \mid^{2}}{35} \label{0++-rate} \\
(10^{3}) B(J/\psi \rightarrow \gamma R(2^{++})) & = &
\left(\frac{m}{1500 MeV}\right) 
\left(\frac{\Gamma_{R\rightarrow gg}}{26 MeV}\right ) 
\frac{x\mid H_{T} \mid^{2}}{34} \label{2++-rate}
\end{eqnarray}
In a similar argument, $\bar{p}p$ annihilations are also considered 
a likely source of glueballs simply because there are so many quarks 
and antiquarks. It is unfortunately difficult to predict rates, and
any signal will be mixed into a background of normal mesons.
Finally double pomeron exchange in central production,
$pp\rightarrow p_{f}{\cal G}p_{S}$ is believed to be a likely source of
glueballs as the pomeron seems to involve glue. In addition to 
glue rich sources, there are glue--poor source such as $2\gamma$
and photoproduction. There is no direct coupling between the
photon and the electrically neutral gluons, so production of glueballs
should be suppressed.

Finally, the glueballs are expected to have a decay pattern
which in some sense is flavour blind. The gluon couples equally to
all flavours of quarks, so the production of $u$, $d$ and $s$ quarks
are expected to be more or less the same. One calculation~\cite{close-88}
yields the values in table~\ref{tab:glueball-decay} as the expected relative 
strengths of for two--pseudoscalar decays. In addition, the $4\pi$ decay 
involving two pairs of $I=0$ $s$--wave dipions, $(\pi\pi)_{s}$ is expected 
to be a significant and large decay mode of a glueball~\cite{close-li}.
This combined with the two--pseudoscalar decays yields the following
expectations.
\begin{table}[htbp]\centering
\begin{tabular}[h]{cccccc}
{\sf Decay} & $\pi\pi$ &  $\bar{K}K$ & $\eta\eta$ & $\eta^{\prime}\eta$ &
$(\pi\pi)_{s}(\pi\pi)_{s}$ \\
{\sf Rate} & $(3)$ & $(4)$ & $(1)$ & $(0)$ & {\sf Large} \\ 
\end{tabular} 
\caption[]{Predicted glueball decay rates.}
\label{tab:glueball-decay}
\end{table}

In particular with the scalar glueball there is the problem of the
nearby scalar mesons. Excluding the $a_{0}(980)$ and $f_{0}(980)$
from consideration, the scalar nonet is presumably made up of the
$f_{0}(1370)$, $a_{0}(1450)$, $K_{0}^{*}(1430)$ and an as yet 
unidentified $f_{0}^{\prime}$ state. The $a_{0}(1450)$ is a recently
identified state with a mass of $1450\pm 40$ and width of $270\pm 40$
observed in $\bar{p}p$ annihilation at rest and decaying into $\eta\pi$,
$\eta^{\prime}\pi$ and $\bar{K}K$~\cite{cbar-a0-eta-pi},
~\cite{cbar-a0-klkl},\cite{cbar-a0-etapr}. Enough of this nonet
is known to allow us to make predictions both on the mass and decay
rates of the missing $f_{0}^{\prime}$ state as a function of the 
nonet mixing angle. If we find additional scalar states, we should be 
able to identify they are pure meson or pure glueball.
\subsection*{The Scalar Sector}
\begin{figure}[htbp]\centering
\vspace{6cm}
\includegraphics{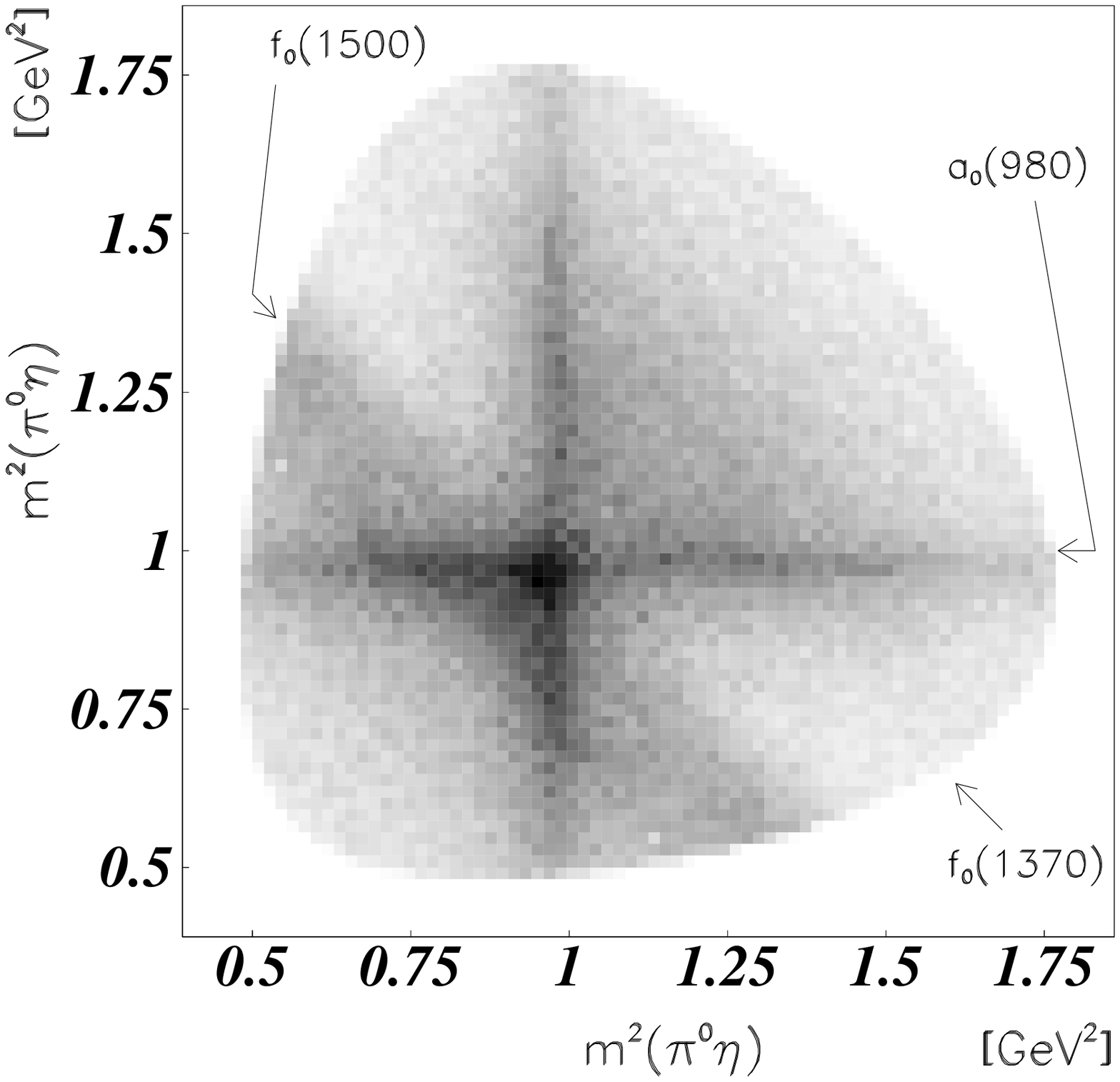}
\includegraphics{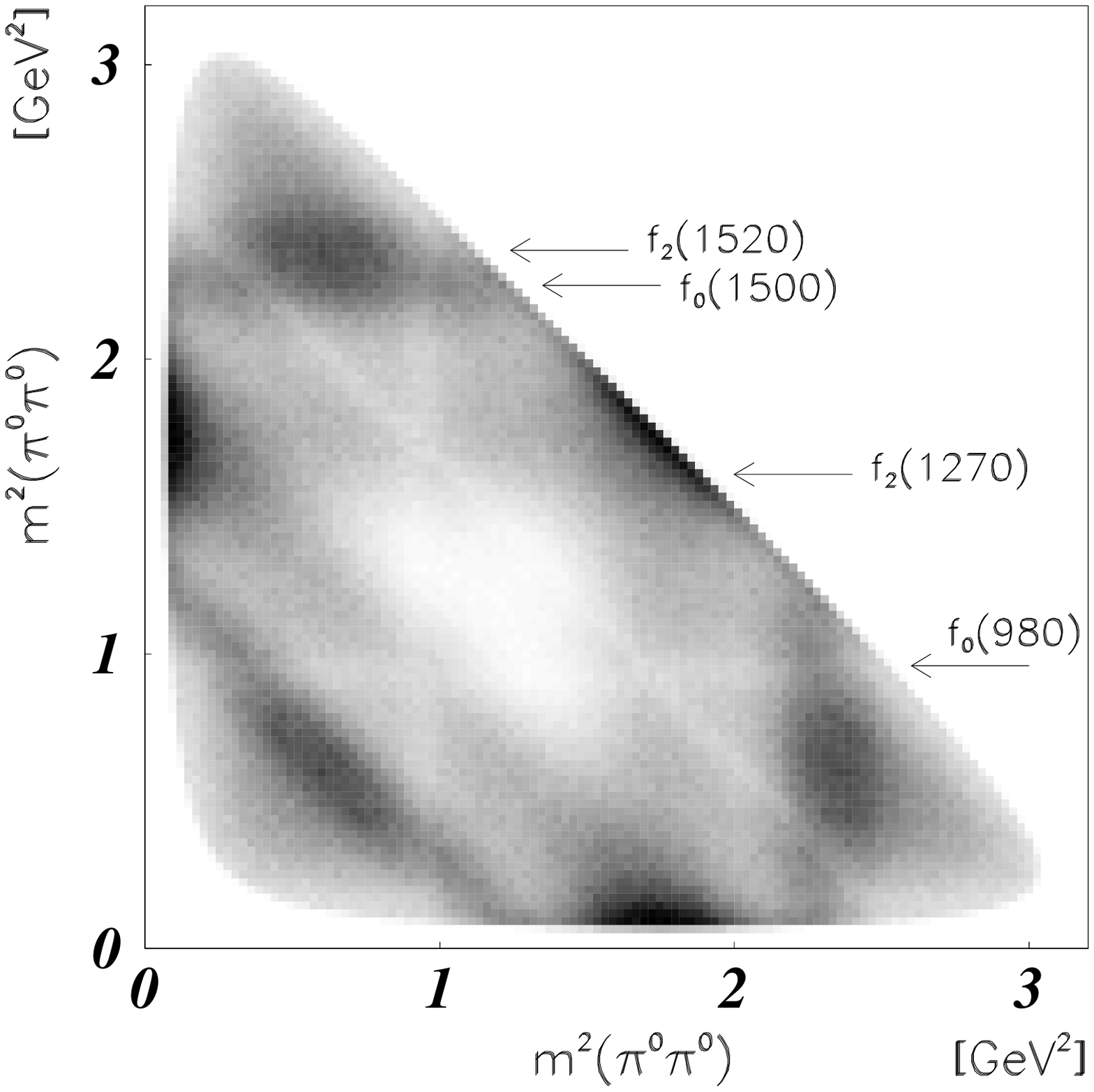}
\caption[]{The Dalitz plots for $\bar{p}p\rightarrow\pi^{\circ}\pi^{\circ}
\pi^{\circ}$ (left) and $\pi^{\circ}\eta\eta$ (right). The $f_{0}(1500)$
is seen clearly as the band near 2.25 GeV$^{2}$ in the $3\pi^{\circ}$ 
Dalitz plot. It is also seen as the lower diagonal band in the 
$\pi^{\circ}\eta\eta$ Dalitz plot.}
\label{fig:dp1}
\end{figure}

\begin{figure}[hbtp]\centering
\vspace{6cm}
\includegraphics{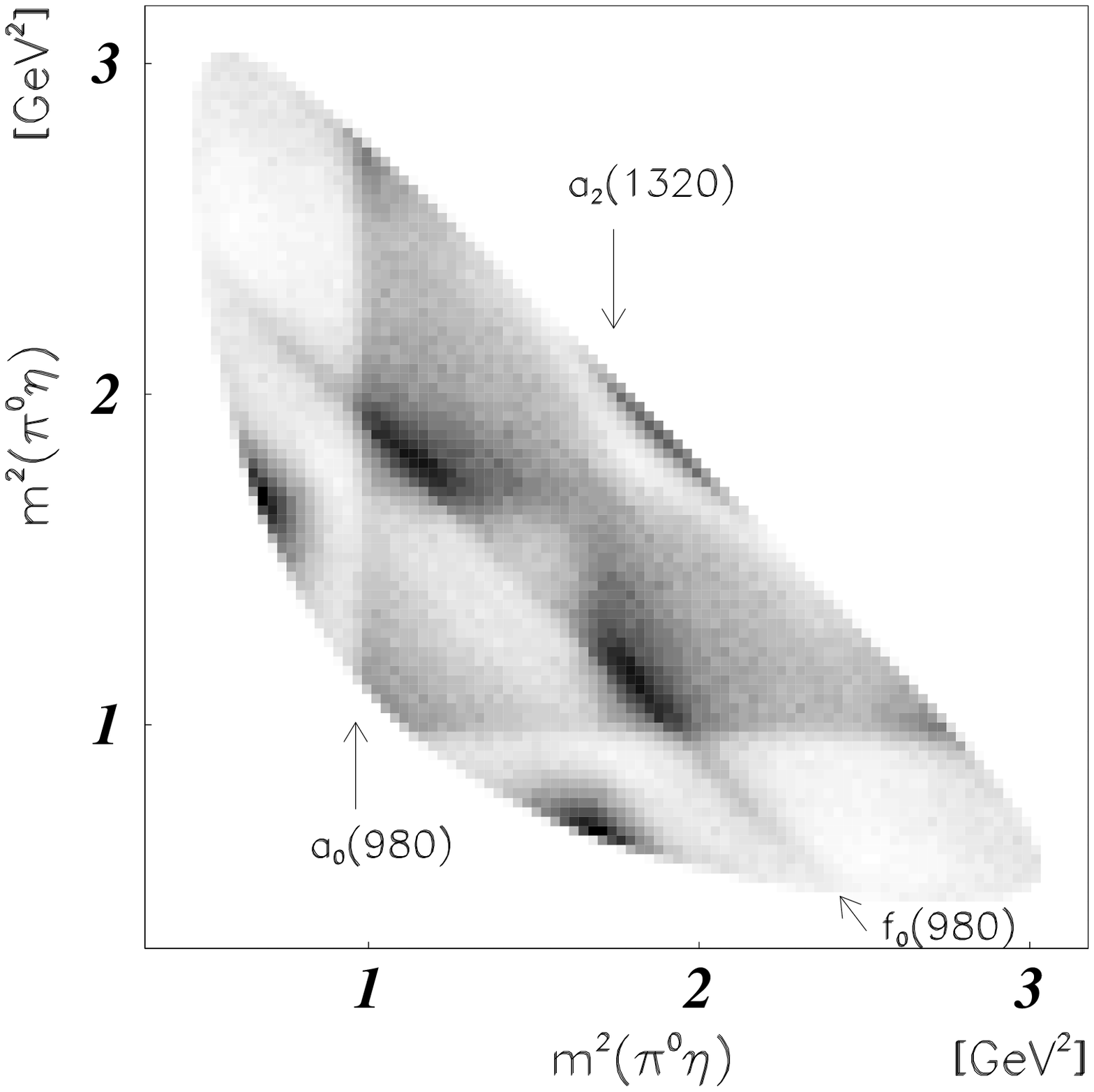}
\includegraphics{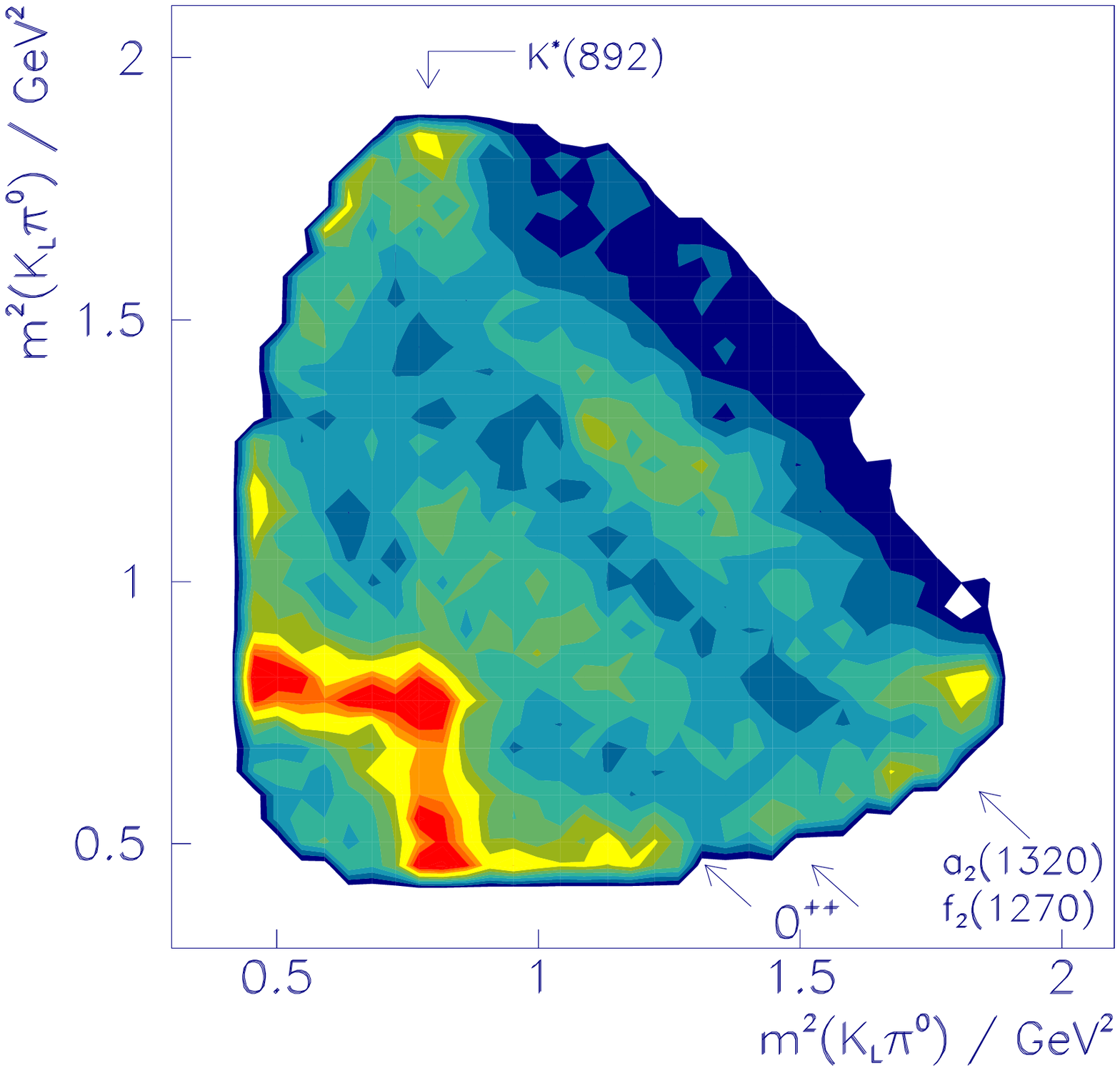}
\caption[]{The Dalitz plots for $\bar{p}p\rightarrow K_{L}K_{L}\pi^{\circ}$ 
(left) and $\pi^{\circ}\pi^{\circ}\eta$ (right). The $f_{0}(1500)$
is seen labeled as the $0^{++}$ band in the $K_{L}K_{L}\pi^{\circ}$ 
Dalitz plot. The $\eta\pi^{\circ}\pi^{\circ}$ Dalitz plot is used
to constrain amplitudes in other channels.}
\label{fig:dp2}
\end{figure}

The Crystal Barrel experiment at LEAR has done a high statistics 
study of $\bar{p}p$ annihilations at rest into both charged and
neutral final states using a nearly $4\pi$ solid angle detector 
for both charged particles and photons. These data have been 
analyzed in a consistent coupled channel analysis. The analysis
is done within the framework of the isobar model using a K--matrix
formulation to maintain unitarity and handle multiple decay modes
of a given meson~\cite{cbar-a0-klkl},~\cite{cbar-couple},~\cite{cbar-etaetapr}.
Dalitz plots for several of these final states
are shown in figures~\ref{fig:dp1} and~\ref{fig:dp2}. In particular,
$\bar{p}p\rightarrow\pi^{\circ}\pi^{\circ}\pi^{\circ}$ ($\sim700,000$ events), 
$\bar{p}p\rightarrow\eta\eta\pi^{\circ}$ ($\sim 200,000$ events) and
$\bar{p}p\rightarrow K_{L}K_{L}\pi^{\circ}$ ($\sim 50,000$ events) 
show a new isoscalar scalar state, 
($(I^{G})J^{PC}=(0^{+})0^{++}$), the $f_{0}(1500)$. Its mass and
width are found to be $m=1500\pm 15$MeV/c$^{2}$ and 
$\Gamma=120\pm 20$MeV/c$^{2}$. In addition to 3--pseudoscalar final states, 
$\bar{p}p\rightarrow\pi^{\circ}\pi^{\circ}\pi^{\circ}\pi^{\circ}\pi^{\circ}$ 
has also been studied~\cite{cbar-5pi0}. These data show evidence for
$f_{0}(1500)\rightarrow (\pi\pi)_{s}(\pi\pi)_{s}$. These consistent
analyses yield the phase space corrected decay rates for the $f_{0}(1500)$ 
as given in table~\ref{tab:0++-decays}.

\begin{figure}[htbp]\centering
\begin{picture}(400,100)
\includegraphics{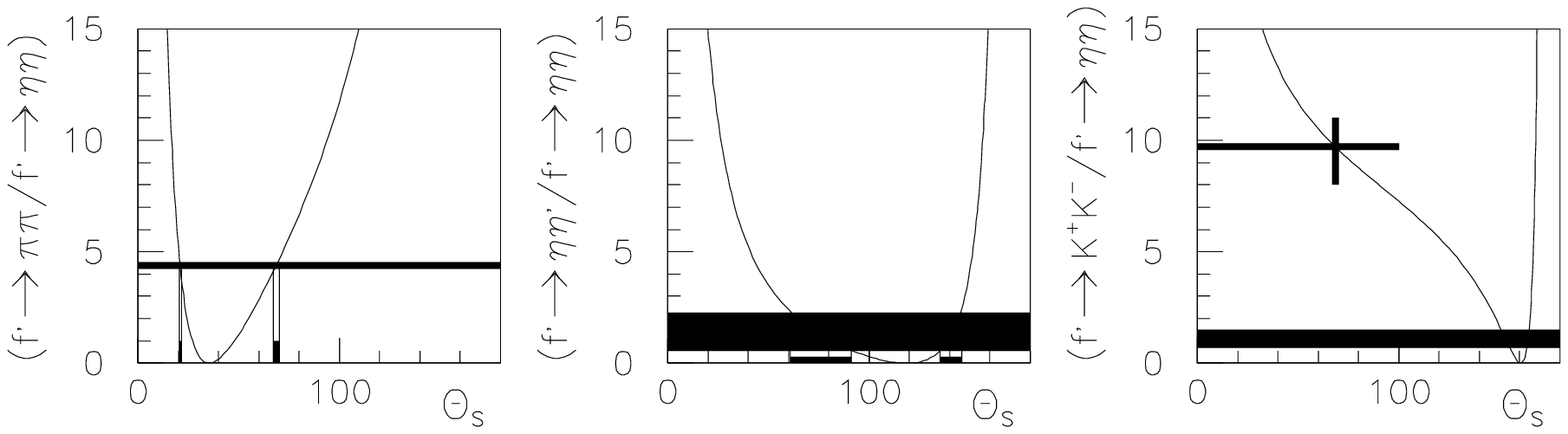}
\put(32,0){\vector(0,1){20}}
\put(58,0){\vector(0,1){20}}
\put(198,0){\vector(0,1){20}}
\put(237,0){\vector(0,1){20}}
\put(340,45){\vector(0,1){20}}
\put(380,84){\vector(-1,0){20}}
\end{picture}
\caption[]{Predicted decay rates of the $f_{0}^{\prime}$ normalized to
$\eta\eta$ as a function of the scalar mixing angle $\theta_{s}$. The
horizontal shaded bands are the Crystal Barrel limits. For $\pi\pi$
and $\eta^{\prime}\eta$ the allowed values of $\theta_{s}$ are indicated
by the arrows. In $\bar{K}K$ the consistent values of $\theta_{s}$ from 
the previous two (upward pointing arrow) are used to predict a ratio. 
This is indicated by the left pointing arrow near a ratio of 9.5.}
\label{fig:ratios}
\end{figure}

Using an SU(3) calculation which accurately predicts the relative
decay rates for the tensor mesons, we can see if the Crystal Barrel
decay rates of the $f_{0}(1500)$ are consistent with the $f_{0}^{\prime}$.
Figure~\ref{fig:ratios} shows the predicted phase space corrected decay 
rates normalized to the $\eta\eta$ decay rate as a function of the
scalar meson mixing angle. The Crystal Barrel data are shown as a 
horizontal band, and the allowed mixing angles are shown as the shaded 
boxes along the mixing angle axes. From the $\pi\pi$ and $\eta^{\prime}\eta$
plots, a consistent value of $\theta_{s}=(68.5\pm1.5)^{\circ}$ is found.
Using this value to predict $\bar{K}K$, we find the region shown by
the shaded cross, and expect a decay ratio of about 10, nearly 9 
times larger than the measured Crystal Barrel Rate. The $f_{0}(1500)$
cannot be the missing $f_{0}^{\prime}$ state. In addition, in
table~\ref{tab:0++-decays} we compare this with the pure glueball.
This is also a poor explanation for this object, the $f_{0}(1500)$
appears to be neither a pure meson nor a pure glueball.
\begin{table}[htbp]\centering
\begin{tabular}[h]{cccccc}
{\sf Decay} & $\pi\pi$ &  $\bar{K}K$ & $\eta\eta$ & $\eta^{\prime}\eta$ &
$ (\pi\pi)_{s}(\pi\pi)_{s}$ \\ \hline
$f_{0}(1500)$ & $(4.39\pm .16)$ & $(1.1\pm .4)$ & $(1)$ & $(1.42\pm .96)$ 
& $(14.9\pm 3.2)$ \\
{\sf GlueBall} & $(3)$ & $(4)$ & $(1)$ & $(0)$ & {\sf Large} \\
$f_{0}^{\prime}$ & $(4.4)$ & $(10)$ & $(1)$ & $(2)$ &  \\
\end{tabular} 
\caption[]{Measured $f_{0}(1500)$ decay rates compared to
a pure glueball and the expected $f_{0}^{\prime}$ state.}
\label{tab:0++-decays}
\end{table}

Finally a third scalar state, the 
$f_{0}(1750)\rightarrow (\pi\pi)_{s}(\pi\pi)_{s}$ is hinted at in the
$\bar{p}p\rightarrow\pi^{\circ}\pi^{\circ}\pi^{\circ}\pi^{\circ}\pi^{\circ}$
final state~\cite{cbar-5pi0}. However, this  state is near the edge
of available phase space and is so far only observed in the one decay 
mode. However as we examine radiative $J/\psi$ decays we find that this
$f_{0}(1750)$ is probably the scalar component of the $f_{J}(1710)$.
A reanalysis of MKIII data~\cite{mk3-bugg} on 
$J\psi\rightarrow\gamma\pi^{+}\pi^{-}\pi^{+}\pi^{-}$ finds two
scalar states both decaying to $(\pi\pi)_{s}(\pi\pi)_{s}$. The $f_{0}(1500)$
and the $f_{0}(1750)$, ($m=1750\pm 15$, $\Gamma=160\pm 40$). In addition 
they find a $2^{++}$ state at a mass of $1620\pm 16$MeV/c$^{2}$ 
and a width $\Gamma=140^{+60}_{-20}$. These latter
two states both come from the $f_{J}(1710)$ region. They find the 
radiative decay rates to the scalars as given in~\ref{mk3-f01500} 
and~\ref{mk3-f01750}.
\begin{eqnarray}
B(J\psi\rightarrow\gamma f_{0}(1500)\rightarrow\gamma 4\pi) 
&=& (5.7\pm 0.8) \times 10^{-4} \label{mk3-f01500} \\
B(J\psi\rightarrow\gamma f_{0}(1750)\rightarrow\gamma 4\pi) 
&=& (9.0\pm 1.3) \times 10^{-4} \label{mk3-f01750}
\end{eqnarray}
Recently, BES has reported both a tensor and scalar state in the
$f_{J}(1710)$ region, $f_{2}(1690)$ ($m=1696\pm 5^{+9}_{-34}$,
$\Gamma=103\pm 8^{+10}_{-31}$) and $f_{0}(1780)$ 
($m=1781\pm 8^{+10}_{-31}$, $\Gamma=85\pm 24^{+22}_{-19}$), both decaying
to $\bar{K}K$~\cite{bes-kk}. These are presumably the same states as
seen in MKIII, and we will adopt the name $f_{0}(1750)$ for the scalar
state. At this conference, BES reported on radiative decays to 
four pions~\cite{bes-zhu}. They find the results;
and~\ref{bes-f01750}.
\begin{eqnarray}
B(J\psi\rightarrow\gamma f_{0}(1500)\rightarrow\gamma 4\pi) 
&=& (4.0\pm 0.6) \times 10^{-4} \label{bes-f01500} \\
B(J\psi\rightarrow\gamma f_{0}(1750)\rightarrow\gamma 4\pi) 
&=& (5.5\pm 0.8) \times 10^{-4} \label{bes-f01750} 
\end{eqnarray}
Taking these together with the fact that $4\pi$ is
on order of 50\% of the decay rate, one finds using
equations~\ref{0++-rate} and~\ref{2++-rate} that
\begin{eqnarray*}
b(f_{0}(1500)\rightarrow gg) & \simeq & 0.5 \sim 0.8 \\
b(f_{0}(1750)\rightarrow gg) & \simeq & 0.5 .
\end{eqnarray*}
Both of which are highly suggestive of a large gluonic content in the 
states~\cite{close-li}.
\begin{figure}\centering
\begin{picture}(400,245)\centering
\includegraphics{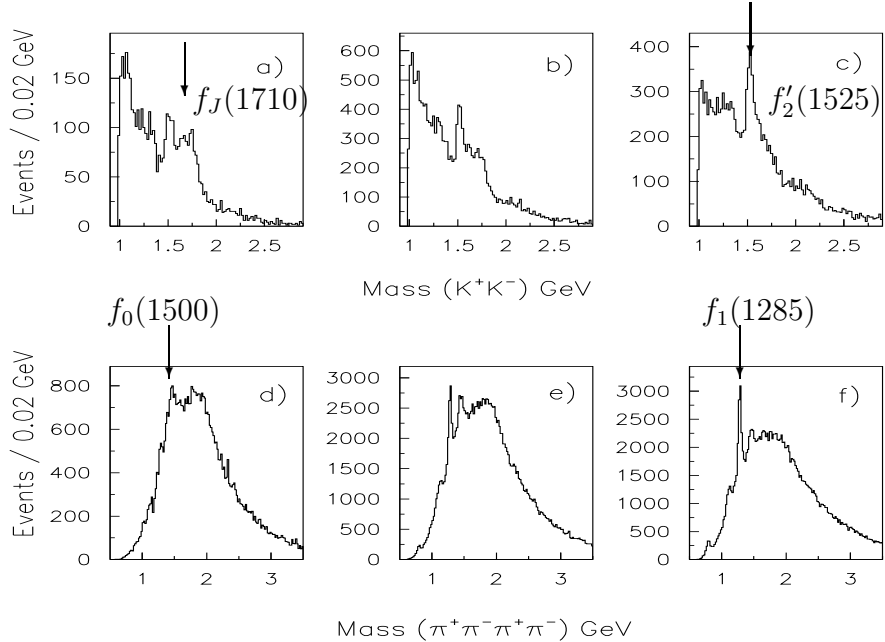}
\put(300,125){\makebox(0,0){$f_{1}(1285)$}}
\put(293,120){\vector(0,-1){20}}
\put(75,125){\makebox(0,0){$f_{0}(1500)$}}
\put(77,120){\vector(0,-1){20}}
\put(325,205){\makebox(0,0){$f_{2}^{\prime}(1525)$}}
\put(297,242){\vector(0,-1){20}}
\put(108,205){\makebox(0,0){$f_{J}(1710)$}}
\put(83,227){\vector(0,-1){20}}
\end{picture}
\caption[]{({\bf a},{\bf d}) $dP_{T}<0.2$, ({\bf b},{\bf e})
$0.2<dP_{T}0.5$ and ({\bf c},{\bf f}) $dP_{T}>0.5$, (see text).}
\label{fig:wa102}
\end{figure}

Finally, we examine data from central production ---
$pp\rightarrow p_{f}(X)p_{s}$. Both the $f_{0}(1500)$
and the $f_{J}(1710)$ have been reported in central production. However
a recent observation made by the WA91 and WA102 collaborations~\cite{wa102}
yields an interesting separation of normal mesons from the exotic candidates.
In central production, double pomeron exchange is believed to be responsible
for formation of gluonic states. However, normal mesons are also produced,
and separation of these has always been difficult. If we consider
the transverse momentum transfer from each proton as $P_{T1}$ and
$P_{T2}$, and the beam axis along $z$,  one can define a variable
\begin{equation} 
dP_{T}=\sqrt{(P_{x1}-P_{x2})^{2}+(P_{y1}-P_{y2})^{2}}.
\label{eq:dpt}
\end{equation}
 In a sense,
this is the difference in transverse momentum between the two pomerons.
When this quantity is small one could visualize the two pomerons 
as traveling together, while when this is large, they are moving apart.
Data for $(X)=(K^{+}K^{-})$ and $(\pi^{+}\pi^{-}\pi^{+}\pi^{-})$
are shown in figure~\ref{fig:wa102}. The interesting feature is 
that for large $dP_{T}$, ({\bf c} and {\bf f}), $K^{+}K^{-}$ shows a clear
signal for the $f_{2}^{\prime}(1525)$ and $4\pi$ shows a clear signal
for the $f_{1}(1285)$ --- both of which are normal mesons. However
in the small $dP_{T}$ region, ({\bf a} and {\bf d}) both of the 
previous states are gone and the $f_{0}(1500)$ and $f_{J}(1710)$
appear. Having the pomerons moving {\it together} enhances the
exotic candidates, while having them move {\it apart} enhances
the normal mesons.

These three production mechanisms taken together lead one to the 
conclusion that both the $f_{0}(1500)$ and the $f_{0}(1750)$ 
have a large gluonic component. Given that these are near 
the scalar mesons, one interpretation is that the pure
$f_{0}$, $f_{0}^{\prime}$ and ${\cal G}$ states have mixed 
and become the observed $f_{0}(1370)$, $f_{0}(1500)$ and $f_{0}(1750)$
states. Two mixing schemes have been proposed to explain this. A
lattice inspired mixing based on computed masses~\cite{ibm-ss-bar}
and a more data inspired approach~\cite{close-li}. Both of
these claim to explain the measured decay rates of the $f_{0}(1500)$
state and are quite similar in their predictions. At this point
it seems we have very strong evidence for a scalar glueball state
mixed into the scalar nonet, and the question is now down to details 
of mixing.
\subsection*{The Tensor Sector}
Assuming we have found the scalar glueball near the lattice predictions,
we also can ask about the next state, the tensor glueball. Here the
information is not nearly as clear, but a new candidate, the $\xi(2230)$ or
$f_{J}(2230)$ has reemerged in recent years due to new measurements
from BES~\cite{bes-p-bar-p}. This state has a mass of 2230 and an 
extremely narrow width of 20. In addition, its decays appear nearly
flavour--blind over $\pi\pi$, $\bar{K}K$ and $\bar{p}p$. It also
appears to have a rather large rate for $J/\psi\rightarrow\gamma\xi$
which would be indicative of a large gluonic component. However it's
spin is currently unclear --- being either $2^{++}$ or $4^{++}$. Even though
it has been reported by BES to decay to $\bar{p}p$, this state has not been 
observed in $\bar{p}p$ direct production. Dave Hertzog~\cite{hertzog}
has given a nice summary of the results and consequences of this
at this conference. The current situation is that these measurements
are just barely compatible, but as the $\bar{p}p$ limits improve,
this may change. In addition, this state has been searched for
and not observed in two--photon production~\cite{galik}. Its nonobservation 
lends support to the glueball interpretation if this state is indeed
a tensor. Finally, BES has recently reported the observation
of $4\pi$ decays of a state $f_{2}(2220)$~\cite{bes-zhu}. They find a spin 2
state at a mass of $m=2220$ and a width of $\Gamma=105^{+70}_{-50}$
decaying to $f_{2}(1270)(\pi\pi)_{s}$. While the width is larger than 
that observed for the $\xi(2230)$, if they are the same state it
is a measurement of the spin. In addition, they find a
radiative rate of
$B(J/\psi\rightarrow\gamma f_{2}\rightarrow\gamma 4\pi) = 
(2.6\times 10^{-4})$.
\section*{Hybrids}
Hybrids are meson--like states to which a valence gluon has been 
added. The quantum numbers of these states are given by those
of the underlying meson plus the quantum numers of the gluon.
These states are predicted in several models, and are expected to
appear in nonets. In the {\em fluxtube} model of Isgur and
Paton~\cite{flux-tube} one expects 8 nonets with quantum numbers
\underline{$0^{+-}$}, $0^{-+}$, $1^{++}$, $1^{+-}$, \underline{$1^{-+}$},
$1^{--}$, \underline{$2^{+-}$} and $2^{-+}$. What makes these states
intriguing is that the \underline{underlined} quantum numbers are not possible
for a $\bar{q}q$ pair. Observation of a state with these quantum
numbers would be a {\em smoking gun} for a non--$\bar{q}q$ state.
The masses of the lightest hybrid mesons are expected to be in the
1700 to 1900 MeV/c$^{2}$ mass range, and an additional signature
is that they are expected to decay into a pair of P and S wave
mesons, {\em e.g.} $f_{1}(1285)\pi$, $a_{2}(1320)\pi$, {\em etc.} 
whereas final states like $\eta\pi$ and $\pi\pi$ should be suppressed. 
In addition,
the hybrids with $\bar{q}q$ quantum numbers might very well
mix with their normal meson counterpart. A recent calculation
of the decays of all mesons and hybrids~\cite{barnes} withing the
framework of the fluxtube model will provide a guide to 
interpreting these states, but we need to observe a state with 
non--$\bar{q}q$ quantum numbers to prove the existence of hybrids.

\begin{figure}[htbp]
\vspace{9cm}
\includegraphics{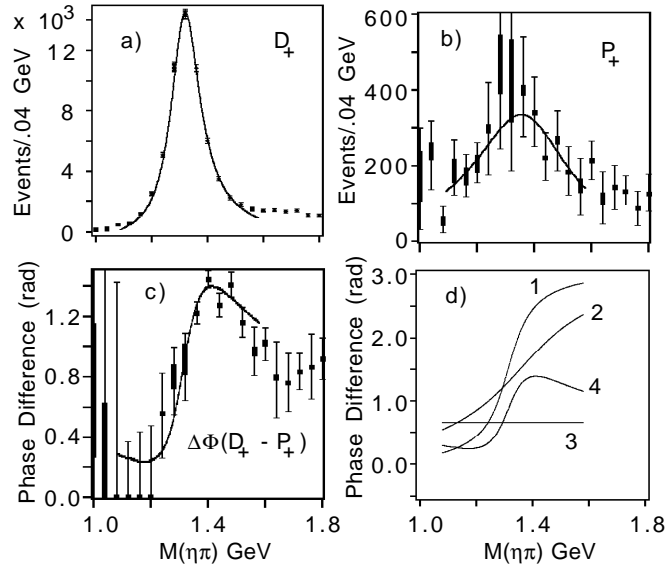}
\caption[]{{\bf a} Data and fit for the $D_{+}$ wave showing a strong 
$a_{2}(1320)$. {\bf b} Data and fits for the $P_{+}$ partial wave. 
{\bf c} Relative phase between the $D_{+}$ and $P_{+}$ partial waves. 
{\bf d} (1) Phase of the $D_{+}$, (2) $P_{+}$, (3) difference and (4) 
production.}
\label{fig-e852}
\end{figure}

At this conference, Brookhaven E852 has presented evidence for
a $1^{-+}$ state decaying into $\eta\pi$~\cite{cason}. This 
state, the $\pi_{1}(1370)$ is observed in the reaction 
$\pi^{-}p\rightarrow\eta\pi^{-}p$ at $18$GeV/c. The state
appears from a detailed partial wave analysis, and is seen 
via its interference with the much stronger $a_{2}(1320)$
state. The data are most easily explained by the introduction
of a resonant state with a mass $m=1370\pm 16^{+50}_{-30}$ and
a width of $\Gamma=385\pm 40^{+65}_{-105}$. In figure~\ref{fig-e852}
are shown the results of their partial wave analysis. The results
yields 8 possible solutions, and the ranges of each value are 
shown as solid vertical bars for the fits. {\bf a} shows the
$D_{+}$ wave which corresponds to the strong $a_{2}(1320)$,
while in {\bf b} is the resulting amplitude of the $P_{+}$
partial wave as a function of $\pi\eta$ mass. The amplitude 
peaks near 1370, and is about 3\% of the strength of the $D_{+}$
wave. In {\bf c} is shown the fit phase difference between the
two waves. In order to explain this, they postulate that a
resonant state, the $\pi_{1}(1370)$ is produced in addition to the
$a_{2}(1320)$. They allow a relative production phase between
these and treat both states as Breit--Wigner resonances. This hypothesis 
is able to reproduce the data, and yields the 4 phases shown in {\bf d}
with the relative production constant over the $\pi\eta$ invariant
mass. Their data agree quite well with earlier data from VES
in the reaction $\pi^{-}N\rightarrow\eta\pi^{-}N$ at $37$Gev/c.
In particular the same phase difference is observed in both data 
sets. 

While there does appear to be something here, it's interpretation is
not so clear. The mass seems too low to be the hybrid meson predicted
by the flux tube model, and the decay mode is also not favored. If it
is a hybrid meson, then we expect it to be a member of a nonet. In 
particular we expect to find both an $\eta_{1}$ and an $\eta^{\prime}_{1}$
state. Depending on their masses, likely decay modes would be either
$\eta\eta^{\prime}$ or $a_{1}(1260)\pi$. The latter would be particularly
difficult to extract from data, but its observation may be quite 
important as establishing this state as a hybrid meson.
\section*{Conclusions}
The scalar meson sector shows strong evidence for a glueball mixed into
the normal mesons. Two states, $f_{0}(1500)$ and $f_{0}(1750)$ have
both been observed in three glue--rich mechanisms. In radiative
$J/\psi$ decays, their large rates are indicative of a large gluonic
content. In central production, both states stand out from normal
mesons via the $dP_{T}$ in equation~\ref{eq:dpt}. 
Finally, in $\bar{p}p$ annihilation,
the relative decay rates of the $f_{0}(1500)$ have been measured and
lead one to the conclusion that the state is neither a pure meson nor a 
pure glueball. The simplest explanation for what we observe is that
the three observed scalars, $f_{0}(1370)$, $f_{0}(1500)$ and $f_{0}(1750)$
are the result of mixing between the pure $f_{0}$ and $f_{0}^{\prime}$
with the scalar glueball ${\cal G}$. In the tensor sector, we have a
rather interesting candidate, the $\xi(2230)$. This state is observed
in radiative $J/\psi$ decays with a rate suggestive of a large gluonic
content, has a very narrow width $\Gamma\sim 20$MeV and seems to have 
flavor blind decays. While more details are needed, this does appear
as our best candidate for the tensor glueball.

Finally, we have evidence for a state with exotic quantum numbers,
$\pi_{1}(1370)$ observed in $\pi^{-}p\rightarrow\eta\pi^{-}p$. The
state is consistent between two different $\pi^{-}N$ experiments,
but its interpretation is murky. Its mass is much lower than one expects 
for a hybrid meson, but observation of one of its partners could 
help clarify this. 
\section*{Acknowledgments}
Discussions with Z. Li, D. Hertzog, and N. Cason are gratefully acknowledged.
This work was supported in part by the U.S. Department of Energy
(contract No. DE-FG02-87ER40315).

\end{document}